\documentclass[12pt]{iopart}
\RequirePackage{graphicx}
\usepackage{latexsym}
\expandafter\let\csname equation*\endcsname\relax
\usepackage{xcolor}
\expandafter\let\csname endequation*\endcsname\relax
\usepackage{amssymb}
\usepackage{amsmath} 
\usepackage{todonotes}

\begin{document}

\title{Holographic vacuum energy regularization and corrected entropy of de Sitter space}

\author{S. Jalalzadeh$^{1,2}$, H. Moradpour$^3$, and H. Tebyanian$^4$ }
\address{$^1$ Departamento de F\'{i}sica, Universidade Federal de Pernambuco, Recife-PE, 50670-901, Brazil }
\address{$^2$ Center for Theoretical Physics, Khazar University, 41 Mehseti Street, Baku, AZ1096, Azerbaijan}
\address{$^3$ Research Institute for Astronomy and Astrophysics of Maragha (RIAAM), P.O. Box 55134-441, Maragheh, Iran}
\address{$^4$ Department of Mathematics, University of York, Heslington, York, YO10 5DD, United Kingdom }
\ead{shahram.jalalzadeh@ufpe.br,  h.moradpour@riaam.ac.ir \& hamid.tebyanian@york.ac.uk}

\vspace{10pt}
\begin{indented}
\item[]March 2024
\end{indented}

\begin{abstract}

We propose that the spectrum of the surface area of the apparent horizon (AH) of de Sitter (dS) spacetime leads to corrected temperature and entropy of the dS spacetime, offering new insights into its thermodynamic properties. This is done by employing the spectrum of the AH radius, acquired from the Wheeler--DeWitt (WDW) equation, together with the Stefan--Boltzmann law, the time-energy uncertainty relation, and the unified first law of thermodynamics.
\end{abstract}

%
%
\submitto{\CQG}
%
%
%

\section{Introduction}\label{sec1man}
In the early 1970s, Jacob Bekenstein revolutionized our understanding of black holes (BHs) by positing that the entropy of a BH is directly proportional to the area of its event horizon \cite{Bekenstein:1973ur}. This groundbreaking hypothesis laid the foundation for the integration of thermodynamic principles into the realm of general relativity. A few years later, in 1975, Stephen Hawking extended this paradigm by introducing the concept of BH evaporation, thereby linking quantum mechanics with the thermodynamics of BHs \cite{Hawking:1975vcx}. Recent observational studies have provided compelling evidence in support of these theoretical frameworks. For instance, gravitational wave observations have been employed to test the area-entropy relationship, yielding results that are consistent with the predictions of Bekenstein and Hawking \cite{Isi:2020tac}.

These empirical validations have invigorated the scientific community's interest in the intricate interplay between thermodynamics, quantum physics, and the geometry of spacetime horizons. Over the years, theoretical physicists have delved deeper into this nexus, uncovering a rich tapestry of relationships that intertwine thermodynamic temperature, quantum field theory, and the geometrical properties of horizons. These discoveries have cemented the role of thermodynamics in the study of gravitational systems and opened up new avenues for exploring the quantum nature of spacetime itself.

{{In quantum gravity, it is possible to derive the thermodynamical characteristics associated with a specific stationary spacetime by employing the technique of Euclidean path integral} \cite{Padmanabhan:2003gd}. The thermal equilibrium state of the system of spacetime, $(\mathcal M,g)$, plus matter field, $\Phi$, for the canonical ensemble is given by the partition function $Z$ as
\begin{equation}
    \label{Re1}
    Z=\int {\mathcal D}g^\text{(E)}{\mathcal D}\Phi
\exp\{I^\text{(E)}[g^\text{(E)},\Phi]\},
\end{equation}
where {\({\mathcal D}g^\text{(E)}\) and \({\mathcal D}\Phi\) represent the path integrals over the Euclidean metric and matter fields, respectively, and} $g^\text{(E)}$ is the Euclidean metric of signature $(++++)$ obtained from the Lorentzian
metric $g$ by the Wick rotation $t\rightarrow -i\tau$, and $I^\text{(E)}[g^\text{(E)},\Phi]$ denotes the Euclidean action functional \footnote{{Because it enhances the convergence of path integrals, a wick rotation of the time axis is frequently employed in quantum field theory. On the other hand, this is not the case in quantum gravity. In general, the integral (\ref{Re1}) is poorly divergent, and the gravitational component of the Euclidean action is unbounded from below. Only partially successful attempts to solve this issue using analytical continuation have been made, and it is still unknown if an integral like (\ref{Re1}) can be defined in a meaningful way \cite{Gibbons:1978ac}.}
}. In order to consider equilibrium states of spacetime with a matter field, it is necessary to impose the periodic boundary condition in imaginary time, $\tau$, for both the metric and the matter field $\Phi$.}

{Note that the fields $g$ and $\Phi$ are not necessarily solutions of classical field equations, however, one can expand these fields as $g=g_\text{cl}+\delta g$ and $\Phi=\Phi_\text{cl}+\delta\Phi$ around the dominant classical solutions when the gravitational and matter fields are weak. As a result, the Euclidean action functional becomes
\begin{equation}
    \label{Re2}
   I^\text{(E)}[g^\text{(E)},\Phi]= I^\text{(E)}[g^\text{(E)}_\text{cl},\Phi_\text{cl}]+ I^\text{(E)}[\delta g^\text{(E)}]+ I^\text{(E)}[\delta\Phi]+\text{higher order terms},
\end{equation}
where {\( g_\text{cl} \) and \( \Phi_\text{cl} \) denote the classical solutions around which the fields are expanded,} $\delta g^\text{(E)}$ and $\delta\Phi$ denote quantum/statistical fluctuations of metric and matter fields, and  $I^\text{(E)}[\delta g^\text{(E)}]$ and $I^\text{(E)}[\delta\Phi]$ are quadratic in fluctuations. }

{Therefore, the partition functional (\ref{Re1}) becomes
\begin{equation}
    \label{Re3}
    \ln Z=\ln Z_\text{cl}+\ln\int \mathcal D[\delta g^\text{(E)}]e^{I^\text{(E)}[\delta g^\text{(E)}]}+\ln\int \mathcal D[\delta\Phi]e^{I^\text{(E)}[\delta\Phi]}+\text{higher order terms},
\end{equation}
where $Z_\text{cl}$ is the classical partition functional. The first term in the above includes only the classical solutions and consequently, the partition function of the thermal equilibrium state of background classical spacetime and matter is defined by $\ln Z_\text{cl}= I^\text{(E)}[g^\text{(E)}_\text{cl},\Phi_\text{cl}]$. Now, utilizing the
Boltzmann’s relation, $S=\ln Z$, one obtains the entropy of the system. This shows that the standard Bekenstein--Hawking entropy (as well as other thermodynamical properties) of an empty background spacetime horizon is encoded in the classical geometry itself. In addition, the correction terms arise due to statistical fluctuations around the thermal equilibrium of the classical geometry and matter field \cite{Saida:2009up, Das:2001ic}. Also, in the presence of a matter field, one should use a proper regularization method to compute the portion of quantum matter field in Eq. (\ref{Re3}). }

{Another approach to acquiring the thermodynamic properties linked to a particular static spacetime is through the utilization of canonical quantization of the relevant gravitational system featuring a horizon, developed by  V.F. Mukhanov \cite{Mukhanov:1986me}. In the case of a Schwarzschild BH, the quantization of the system results in a discrete spectrum of horizon areas. }
Bekenstein pioneered the concept of BH horizon area quantization, proposing that the area, $A_{\text{BH}}$, could be represented by discrete eigenvalues \cite{Bekenstein:1974jk}:
\begin{equation}
    \label{H1}
    A_\text{BH}(n)=\gamma L^2_\text{P}{\,}n,~~~~n=\{\text{large integers}\},
\end{equation}
where $\gamma$ is a dimensionless constant of order one, depending on the details of the quantization method \cite{Ashtekar:1997yu}, and $L_\text{P}=\sqrt{G}$ is the Planck length. Subsequently, numerous studies (see, for example, \cite{doi:10.1142/8540} and the references therein) have been integrated into it, all of which have played a role in bolstering the accuracy of the mass spectrum (\ref{H1}).

It can be hypothesized that a BH emits Hawking radiation during its spontaneous transition from state $n+1$ to the nearest lower state level, specifically, $n$. Supposing the emitted thermal radiation's frequency is denoted as $\omega_0$, it follows that
\begin{equation}
   \label{H5}
   \omega_0=M^{(n+1)}-M^{(n)}=\frac{\gamma M_\text{P}}{\sqrt{2n}}+\text{higher order terms}=\frac{\gamma^2M^2_\text{P}}{2M}+\text{higher order terms}.
\end{equation}
According to Mukhanov \cite{Mukhanov:1986me} this equation bridges the gap between quantum mechanics and classical thermodynamics in the context of BHs.

{The argument employed by Mukhanov to derive the thermodynamical characteristics of a BH, as detailed in the reference \cite{Mukhanov:1986me}, can be elucidated in the following manner: Initially, he assumed that the spectrum of a BH adheres to the Eq. (\ref{H1}). Subsequently, in his analysis to determine the rate of Hawking radiation, he proceeded to calculate the BH's lifetime while it is in a specific quantum state represented by the quantum number $n$, as expressed in Eq. (\ref{H5}). Lastly, by incorporating the Stefan--Boltzmann law in conjunction with the first law of thermodynamics and Mukhanov's argument, as elaborated in Refs. \cite{Xiang:2004sg, Jalalzadeh:2021gtq, Jalalzadeh:2022rxx}, one is able to deduce the temperature and entropy of a Schwarzschild BH.}

{The advantage of this method of calculating entropy to Euclidean path integral is its simplicity in calculating the higher-order correction terms: a Taylor expansion of (\ref{H5}) at the desired order, gives us the corresponding order of correction term of the entropy as well as the temperature of the BH. In addition, sometimes,  we just know the modified WDW equation, and the corresponding path integral is not developed enough. For example, in Ref. \cite{Jalalzadeh:2021gtq}, the authors developed a fractional WDW equation of a Schwarzschild BH. Also, a quantum deformed (through quantum groups) extension of the WDW equation is represented in \cite{Jalalzadeh:2022rxx} for a Schwarzschild BH. In this method, the corresponding extension of (\ref{H5}) immediately gives us the corresponding entropy of the model under investigation.} 

{In this article, we extend the aforementioned approach to a closed Friedmann--Lemaître--Robertson--Walker (FLRW) universe with a positive cosmological constant, in which the matter content is radiation. To achieve this objective, our first task is to address the WDW equation of the model universe to acquire the appropriate wavefunction. On the other hand, in order to acquire a finite value for the spectrum of the cosmological constant, it is imperative to employ an appropriate regularization technique that ensures a finite value for the vacuum energy associated with the electromagnetic field. As a result of the holographic regularization method, a discrete spectrum emerges for the dS AH, mirroring the characteristics of the spectrum observed for a BH as delineated in Eq. (\ref{H1}). Note that one can use other regularization methods \cite{Jalalzadeh:2022bgz, Ford:1976fn} to keep vacuum energy finite. However, the benefit of holographic regularization is that the classical universe emerged from the quantum region as a universe without the horizon, flatness, singularity, and cosmic coincidence problems \cite{Jalalzadeh:2022dlj}.}

The present paper is organized in the following manner: In Section \ref{sec2man}, we delve into a brief discussion regarding the WDW equation that governs the behaviour of a closed FLRW universe in the presence of radiation as a matter field. Upon establishing the fact that the dS AH radius undergoes quantization, we proceed to Section \ref{sec3man}, where we employ semiclassical arguments, to obtain the classical emergent dS universe. In section \ref{Sentropy} the Stefan--Boltzmann law, the time-energy uncertainty relation, and the unified first law of thermodynamics utilized to derive the corrected temperature and entropy of dS space emerged from quantum cosmology. Lastly, Section \ref{sec4man} encompasses the conclusion that further elucidates the findings and implications presented throughout this paper.


\section{Quantum cosmology with holographic regularized vacuum energy}\label{sec2man}

Consider a background minisuperspace, a spatially closed ($k=1$), homogeneous, and isotropic metric. The line element is
\begin{equation}\label{1-1}
ds^2=-N^2(t)dt^2+a^2(t)h_{ij}dx^idx^j,
\end{equation}
in which $a(t)$ is the scale factor, $N(t)$ is the lapse function, and $h_{ij}dx^idx^j=d\chi^2+\sin^2(\chi)d\Omega_{(2)}^2$ is the line element on the unit 3-sphere, $\mathbb S^3$.

{Consider the action functional for gravity and a minimally coupled electromagnetic field $F_{\mu\nu}$ with the standard form
\begin{equation}\label{1-1aaa}
I=\frac{1}{16\pi G}\displaystyle\int_\mathcal{M}(R-2\Lambda)\sqrt{-g}d^4x
-\frac{1}{4}\displaystyle\int_\mathcal{M}F_{\mu\nu}F^{\mu\nu}\sqrt{-g}d^4x+S_\text{GHY},
\end{equation}
where $S_\text{GHY}$ is the Gibbons--Hawking--York boundary term, $g$ is the determinant of the metric (\ref{1-1}), and $\Lambda$ is the cosmological constant. 
By Fourier expansions \cite{Jalalzadeh:2022bgz} of electromagnetic 4-vector fields $A_\mu$ 
\begin{equation}\label{Re31}
\begin{split}
A_0(x,t)&=\sum_{j=0}^\infty\sum_{l=0}^j\sum_{m=-l}^{l}g_{jlm}(t)Y_{jlm}(x),    \\
A_k(x,t)&=\sum_{B=0}^2\sum_{j=j_\text{min}}^\infty\sum_{l=l_\text{min}}^j\sum_{m=-l}^l f^{jlm}_{(B)}(t)Y^{jlm}_{(B)k},~~~~k=1,2,3,
\end{split}
\end{equation}
in terms of the scalar hyperspherical harmonics, $Y_{jlm}(x)$, and the vector hyperspherical harmonics, $Y^{jlm}_{(B)k}$}
into the Maxwell action functional in the background (\ref{1-1}), {and choosing the Coulomb-type (or radiation) gauge condition
\begin{equation}\label{Re6}
A_0=0,\hspace{.5cm}^{(3)}\nabla^iA_i=0,
\end{equation}}
we arrive at the Arnowitt--Deser--Misner (ADM) Lagrangian and Hamiltonian (which may be found in \cite{Jalalzadeh:2022bgz} for computational details), respectively
\begin{equation}
    \label{1-17a}
    L_\text{ADM}=\frac{3\pi}{4G}\Big(-\frac{a\dot a^2}{N}+Na-\frac{\Lambda}{3}a^3\Big)+
    \frac{1}{2}\sum_{B=1}^2\sum_{j=1}^\infty\sum_{l=1}^j\sum_{m=-l}^l\Big(\frac{a}{{N}}\dot f^2_{(B)J}-\frac{N}{a}(j+1)^2f^2_{(B)J}\Big),
\end{equation}

\begin{multline}
    \label{1-22}
    H_\text{ADM}=N\Big\{-\frac{G}{3\pi a}\Pi_a^2-\frac{3\pi}{4G}a+\frac{\pi\Lambda}{4G}a^3+\\
    \frac{1}{2a}\sum_{B=1}^2\sum_{j=j_\text{min}}^\infty\sum_{l=l_\text{min}}^j\sum_{m=-l}^l\left(\Pi_{(B)jlm}^2+(j+1)^2f_{(B)jlm}^2\right) \Big\},
\end{multline}
where the expansion coefficients of $A_i$, $f^{jlm}_{(B)}(t)$, depend only on the cosmic time $t$. Also, $\Pi_a=-\frac{3\pi}{2G}\frac{a\dot a}{N}$ and $\Pi_{(B)J}=\frac{a}{N}\dot f_{(B)J}$ (we use compact notation $J=\{j,l,m\}$) are the conjugate momenta of the scale factor, $a$, and $f_{(B)J}$, respectively.

The ADM action of the model is given by
\begin{equation}
    I_\text{ADM}=\frac{1}{16\pi G}\int_{t_i}^{t_f}\Big\{\dot a\Pi_a+\sum_{B,J}\dot f_{(B)J}\Pi_{(B)J} -N \mathcal H\Big\}dt,
\end{equation}
where $ \mathcal H$ is the super-Hamiltonian, given by
\begin{equation}\label{1-23}
  \mathcal H=-\frac{G}{3\pi a}\Pi_a^2-\frac{3\pi}{4G}a+\frac{\pi\Lambda}{4G}a^3+   \frac{1}{2a}\sum_{B,J}\left(\Pi_{(B)J}^2+(j+1)^2f_{(B)J}^2\right).
\end{equation}

Therefore, the corresponding phase space is the cotangent bundle
\begin{equation}
    \Omega=\{a, f_{(B)J}, N, \Pi_a, \Pi_{(B)J}, \Pi_N  \}.
\end{equation}
The ADM Lagrangian of the model (as all ADM Lagrangians in GR) is singular because the conjugate momenta $\Pi_N$ weakly vanishes
\begin{equation}
    \Pi_N=\frac{\partial L_\text{ADM}}{\partial\dot N}\approx0,
\end{equation}
meaning that the Legendre transformation
\begin{equation}
 \{a, f_{(B)J}, N, \dot a, \dot f_{(B)J}, \dot N  \}\rightarrow   \{a, f_{(B)J}, N, \Pi_a, \Pi_{(B)J}, \Pi_N  \},
\end{equation}
is a surjective map and not an invertible map. Thus, we are working with a constrained Hamiltonian system since the Legendre transform is non-invertible. According to Dirac constraint theory, the fact that $\Pi_N$ is weakly vanish means that N, the laps function, is a freely selectable and physically irrelevant variable. Thus, the primary Hamiltonian of the model is $H_{Pr}=H_\text{ADM}+\lambda\Pi_N$, where $\lambda$ is a Lagrange multiplier. Now, we have
\begin{equation}
    \dot\Pi_N=\{\Pi_N, H_{Pr}\}=\mathcal H.
    \end{equation}
This demonstrates that the primary constraint is not preserved, necessitating the imposition of the following secondary \cite{doi:10.1142/8540} constraint
\begin{equation}
    \label{HRRR}
    \mathcal H\approx0.
\end{equation}

{Utilizing the quantization map in the configuration space
\begin{equation}\label{quan}
\begin{split}
(\hat a, \hat\Pi_a)&= (a,-i\frac{\partial}{\partial a}),\\ (\hat f_{(B)J},\hat\Pi_{(B)J})&= (f_{(B)J},-i\frac{\partial}{\partial {f_{(B),J}}}).
\end{split}
\end{equation}
the super-Hamiltonian constraint (\ref{HRRR}) transforms to the WDW equation \cite{Jalalzadeh:2022bgz,Jalalzadeh:2022dlj}, where the super-Hamiltonian annihilate wavefunction
\begin{multline}
    \label{2-29}
    \Bigg\{\frac{1}{3\pi a M_\text{P}^2}a^q\frac{d}{da}\left(a^{-q}\frac{d}{da}\right)+\frac{3\pi M_\text{P}^2}{4}a-
    \frac{1}{2a}\sum_{BJ}\Big(-\partial_{(B)J}^2\\+(j+1)^2f_{(B)J}^2\Big)\Bigg\}\Psi(a,f_{(B)J})=0,
\end{multline}
where  $q$ is the Hartle--Hawking--Verlinde \cite{PhysRevD.28.2960,PhysRevD.33.3560,PhysRevD.37.888} factor ordering parameter.}

{ Note that the electromagnetic part of the above super-Hamiltonian,
\begin{equation}\label{2-31}
    \mathcal H_m=\frac{1}{2a}\sum_{BJ}\left(-\partial_{(B)J}^2+(j+1)^2f_{(B)J}^2\right),
\end{equation}
represents the contribution of an infinite number of harmonic oscillators with eigenfrequencies $\omega_j=(j+1)/a$. }

{If we rewrite the matter part of the super-Hamiltonian, instead of $(f_{(B)J},\Pi_{(B)J})$, in terms of the creation, $\hat C^\dagger_{(B)J}$, and annihilation, $\hat C^\dagger_{(B)J}$, operators
\begin{equation}
   \begin{split}\label{1-24}
   \hat C_{(B)J}:=\frac{1}{\sqrt{2(j+1)}}\left(\partial_{(B)J}+(j+1)f_{(B)J}\right),\\ 
  \hat C^\dagger_{(B)J}:=\frac{1}{\sqrt{2(j+1)}}\left(-\partial_{(B)J}+(j+1)f_{(B)J}\right),\\
  \left[\hat C_{(B)J},\hat C^\dagger_{(B')J'}\right]=i\delta_{BB'}\delta_{JJ'},
\end{split} 
\end{equation}
then, (\ref{2-31}) will be
\begin{equation}\label{2-33}
    \mathcal H_m=\frac{1}{a}\displaystyle\sum_{B=1}^2\sum_{j=1}^\infty\sum_{l=1}^j\sum_{m=-l}^l(j+1)\left(\hat N_{(B)jlm}+\frac{1}{2}\right),
\end{equation}
where $\hat N_{(B)J}:=\hat C^\dagger_{(B)J}\hat C_{(B)J}$ are number operators.
According to the above Hamiltonian, the vacuum energy of the electromagnetic radiation is 
\begin{equation}\label{EE1}
    \mathcal H_m^\text{(vacuum)}=\frac{1}{a}\sum_{j=1}^\infty j(j^2-1).
\end{equation}
 All vacuum modes contribute to the zero-point energy, causing the sum in the above relation to be divergent. Therefore, we must utilize a regularization method to reduce it to a finite value. Inserting (\ref{EE1}) into the WDW equation (\ref{2-29}) gives us the following WDW equation in the presence of the electromagnetic vacuum energy}
\begin{equation}
    \label{2-29bbb}
    \Bigg\{-a^q\frac{d}{da}\left(a^{-q}\frac{d}{da}\right)+\left(\frac{3\pi}{2L_P^2}\right)^2a^2\left(1-\frac{a^2}{L_\Lambda^2}\right)
   - {3\pi aM_\text{P}^2}  \mathcal H_m^\text{(vacuum)}\Bigg\}\psi(a)=0,
\end{equation}

 The contribution of vacuum energy is boundless, posing a significant challenge to comprehending and regulating it effectively in the WDW equation. However, one promising approach to regularize this energy is through holographic regularization, as recently formulated in Ref. \cite{Jalalzadeh:2022dlj}. This novel technique, in summary, employs the Cohen--Kaplan--Nelson holographic bound \cite{holography-appl}, which suggests that the UV cutoff of an effective quantum field theory in a box of size $L$ should be low enough that states of characteristic energy density are not BHs.  Employing this bound the authors of \cite{Jalalzadeh:2022dlj} showed that there exist a $j_\text{max}$ in which 
\begin{equation}\label{abb0}
 \sum_{j=1}^{j_\text{max}} j(j^2-1)\simeq\frac{j_\text{max}^4}{4}=\frac{1}{4}\left(\frac{L_\Lambda}{L_\text{P}}\right)^2,
\end{equation}
 as the cutoff on the wave number of the virtual photons. As per the findings in \cite{Jalalzadeh:2022dlj}, holographic vacuum energy regularization leads to the inception of a classical universe at the Planck time, with a much larger scale factor than that of the Planck length. Additionally, the emergence of such a universe can take place without any singularity, horizon, or flatness issues. Notably, it proposes a fresh, non-anthropic solution to the cosmic coincidence problem, which could potentially aid in eliminating fine-tuning.

Inserting (\ref{EE1}) into the WDW equation (\ref{2-29bbb}) conducts us to the scale factor part of the WDW equation, which is expressed as
\begin{equation}\label{2-39}
 -a^q\frac{d}{da}\left(a^{-q}\frac{d\psi}{da}\right)+\left(\frac{3\pi}{2L_P^2}\right)^2a^2\left(1-\frac{a^2}{L_\Lambda^2}\right)\psi(a)=\frac{L_\Lambda^2}{L_P^4}\psi(a).
\end{equation}
\begin{figure}[h!]
  \centering
\includegraphics[width=8cm]{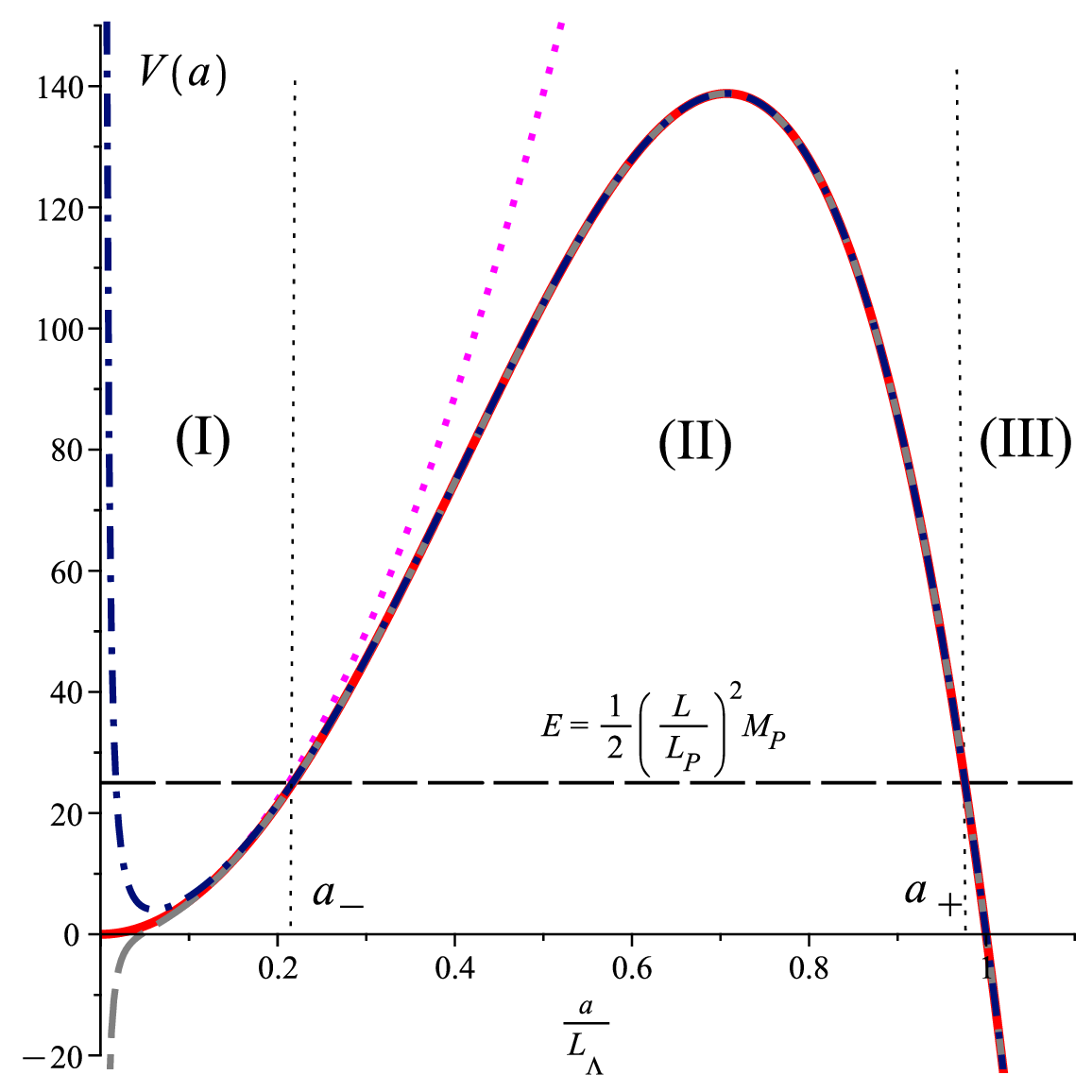}
  \caption{\small The `particle's' potential energy as a function of scale factor in the WDW Eq. (\ref{R1}). The plot of $V(a)$ for $q>0$ or $q<-2$ is shown by the dot-line. The potential energy plot for $-2<q<0$ is shown by the dashed line, while the $V(a)$ plot for $q=0$ is represented by the red line. The dotted line shows the harmonic potential term of $V(a)$.  }\label{fig}
\end{figure}

{Let us redefine the wavefunction as $\pi=a^\frac{q}{2}\Phi(a)$. Then WDW equation (\ref{2-39}) will reduce to 
\begin{equation}
    \label{R1}
   \frac{1}{2M_\text{P}} \frac{d^2\Phi}{da^2}+(E-V(a))\Phi(a)=0,
\end{equation}
where realizes a Schrodinger equation of a `particle' with Planck mass, and the following potential energy, and energy, respectively
\begin{equation}
    \label{R2}
    V(a)=\frac{1}{2M_\text{P}}\left\{\frac{\frac{q}{2}(\frac{q}{2}+1)}{a^2}+\left(\frac{3\pi}{2L^2_\text{P}}\right)^2a^2\left(1-\frac{a^2}{L_\Lambda^2} \right)\right\},
\end{equation}
\begin{equation}
    \label{R3}
    E=\frac{1}{2}\left(\frac{L_\Lambda}{L_\text{P}}\right)^2M_\text{P}.
\end{equation}
The particle energy, $E$, is always less than the maximum of the potential barrier
\begin{equation}
    \label{R4}
    V_\text{max}=\frac{9\pi^2}{32}\left(\frac{L_\Lambda}{L_\text{P}}\right)^2M_\text{P}=\left(\frac{3\pi}{4}\right)^2E.
\end{equation}
In addition, the classical turning points, see Fig. \ref{fig}, are given by
\begin{equation}
    \label{R5}
    a_\pm=\sqrt{\frac{1}{2}\pm\frac{1}{2}\sqrt{1-\frac{16}{9\pi^2}}}L_\Lambda.
\end{equation}
}

{It is straightforward to verify that at the turning point $a_-$, the ratio of the anharmonic term to the harmonic term of the potential is equal to $a_-^2/L_\Lambda^2\simeq0.05$. This suggests that the anharmonic part of the potential energy in (\ref{R4}) can be ignored in the area (I) of Fig. \ref{fig} and in the semi-classical approximation. This is seen in Fig. \ref{fig}, where the potential energy coincides with the quadratic potential (indicated with dots) up to the classical turning point for all values of the ordering parameter $q$.
}

{Ignoring the anharmonic term of potential in (\ref{R2}) in the region (I), the solution of the WDW equation (\ref{R1}) is a confluent hypergeometric (Kummer) function. Thus,
\begin{equation}
    \label{R6}
    \psi(a)=a^\frac{q}{2}\Phi(a)=Ce^{-\frac{3\pi a^2}{4L_P^2}}a^{q+1}~
    _1F_1\left(-\frac{1}{6\pi}\left( \frac{L_\Lambda}{L_P}\right)^2;\frac{q+3}{2};\frac{3\pi a^2}{2L_P^2}\right),
\end{equation}
where $C$ is a constant of normalization.
}

In order for the wavefunction to be square-integrable, it is imperative that the hypergeometric series $_1F_1(\alpha;\beta;\xi)$ be a finite polynomial of degree $n$. This can be accomplished if and only if a non-negative integer $n$ exists, such as $\alpha=-n$. In consequence, one arrives at the quantization of the dS AH radius
\begin{equation}
    \label{new2-3}
    L_\Lambda=\sqrt{6\pi \left(n+\frac{3+q}{4}\right)}L_\text{P}\simeq\sqrt{6\pi n}L_\text{P}.
\end{equation}
Since we are interested in semiclassical approximation ($n\gg1$), we omitted $(3+q)/4$. Refs. \cite{Xiang:2004pf,Jia:2011zzb} yields a similar conclusion for quantization of dS radius; however, there, the coefficient of $n$ is $2$ rather than $6\pi$.

The energy scale of inflation is typically expected to be around $10^{-6}M_\text{P}$. Therefore, $L_\Lambda\simeq 10^{6}L_\text{P}$, which shows that $n\simeq10^{11}$.
 
\section{Emerged de Sitter universe}\label{sec3man}

In the present paper, we focus on a semiclassical approximation, disregarding operator ordering. Consequently, we set the ordering parameter $q$ to zero. The WKB, Linde--Vilenkin or Hawking no-boundary, wavefunction of the one-dimensional WDW equation mentioned above for $a\geq a_+$, (region III of Fig. \ref{fig}), is
\begin{equation}
    \label{psiWKB}
    \psi(a)_\pm\propto\frac{1}{\sqrt{\Pi_a}}\exp\left\{\pm i\int_{a_+}^a\Pi_ada\mp i\frac{\pi}{4}\right\},
\end{equation}
where $\Pi_a$ is the classical conjugate momenta of the scale factor, obtained from the classical equation
\begin{equation}\label{RRRp}
    \Pi_a=-\sqrt{2M_\text{P}(E-V)}\simeq -\frac{L_\Lambda}{L_\text{P}^2}\sqrt{1+\left(\frac{3\pi}{2L_\Lambda^2}\right)^2a^4}.
\end{equation}
Note that in this relation we ignored the factor ordering and the harmonic potential terms in potential (\ref{R4}) for $a\gg a_+$. 

{For the leading order in the WKB approximation, it is easy to verify that the wavefunctions} (\ref{psiWKB}) are also the eigenvectors of the operator $\Pi_a$ \cite{Vilenkin:1994rn}
\begin{equation}
    \hat\Pi_a\psi_\pm(a)=\pm \Pi_a\psi_\pm(a),
\end{equation}
Therefore, the wavefunction described above exhibits an oscillatory pattern, indicating an infinitely expanding, $\psi_-(a)$, (also a contracting, $\psi_+(a)$) and the flat universe. The definition of the conjugate momenta of the scale factor $\Pi_a=-\frac{3\pi}{2G}\frac{a\dot a}{N}$ in (\ref{1-22}), and (\ref{RRRp}) result 
\begin{equation}\label{wave1}
    a^2\dot a^2=\left(\frac{2L_\Lambda}{3\pi} \right)^2\left\{1+\left(\frac{3\pi}{2L_\Lambda}\right)^2a^4 \right\},
\end{equation}
leading to the Friedmann equation
\begin{equation}\label{wave1aa}
    H^2 =\frac{4}{3\pi^2\Lambda}\frac{1}{a^4}+\frac{\Lambda}{3},
\end{equation}
where we used the definition of $L_\Lambda=\sqrt{3/\Lambda}$.

It is worth noting that the Friedmann equation (\ref{wave1aa}) shows that the classical flat universe with radiation and cosmological constant emerged after the quantum tunnelling of the universe for $a>a_+$. On the other hand, the same equation shows that the ratio of the radiation density, $\rho_\gamma=\frac{4}{3\pi^2\Lambda}\frac{1}{a^4}$, to the density cosmological constant, $\rho_\Lambda=\frac{\Lambda}{8\pi G}$ at $a_+$ is
\begin{equation}
    \frac{\rho_\gamma}{\rho_\Lambda}\Big|_{a=a_+}=\frac{8}{9\pi^2}\frac{1}{1+\sqrt{1-\frac{16}{9\pi^2}}}\simeq0.05.
\end{equation}
This means that the emerged universe is already cosmological constant dominated just after tunnelling. Therefore, for $a>a_+$ (\ref{wave1aa}) gives us the dS spacetime $a=a_0\exp\left(\frac{t}{L_\Lambda}\right)$
with the Hubble constant 
\begin{equation}\label{wave6}
   H =\frac{1}{L_\Lambda}=\sqrt{\frac{\Lambda}{3}}.
\end{equation}
The aforementioned Hubble parameter is responsible for the expansion of dS at a later stage.

Note that in quantum cosmology, usually the assumption is tunnelling of the Universe from `nothing' into dS universe \cite{Linde:1983mx, Vilenkin:1985md, Vilenkin:2018dch, Jalalzadeh:2022uhl}. Therefore, in these approaches, we have only regions 2 and 3 of Fig. \ref{fig}. However, in the model studied here, we have a finite vacuum energy of the electromagnetic field, which is proportional to the square of the dS radius. This energy generates region I of Fig. \ref{fig}. 

Note that in quantum cosmology, the quantum universe is often assumed to tunnel from ``nothing'' into the dS space. As a result, only regions II and III of Fig. \ref{fig} are present in these models. On the other hand, the electromagnetic field in the model under study has a finite vacuum energy that is proportional to the square of the dS radius. The energy in question produces region I of Fig. \ref{fig} and is responsible for the quantisation of $L_\Lambda$ in (\ref{new2-3}).

{At this juncture, we are poised to synthesize our findings. This synthesis will serve to elucidate our route in subsequent sections aimed at deriving the temperature and entropy of dS spacetime. Our observations indicate that the model universe tunnels through a potential barrier and emerges in a dS universe, characterized by a quantized cosmological constant. Consequently, as delineated in Eq. (\ref{new2-3}), the resultant system resembles a simplistic model akin to the hydrogen atom. In the hydrogen atom, the interaction with electromagnetic radiation leads to the spontaneous emission and absorption of electromagnetic radiation. This process entails emitted radiation energy equating the energy differential between the atom's energy levels. It is noteworthy that the atomic spectrum exhibits non-negligible width \cite{Sakurai:2011zz} and the emitted radiation adheres to the Stefan--Boltzmann law when in a state of equilibrium. }

{Similarly, we propose that the energy of emitted Hawking radiation from the apparent horizon of dS space is equal to the Misner--Sharp--Hernandez (MSH) energy difference of the simple quantized dS system between two states $n+1$ and $n$. This, in combination with the energy-time uncertainty relation, leads us to calculate the time variation of MSH energy of dS spacetime due to thermal radiation. Assuming the system is in thermodynamic equilibrium, the emitted thermal radiation follows the Stefan-Boltzmann law, similar to the hydrogen atom example. This assumption allows us to determine the dS temperature. Finally, by using the unified first law of thermodynamics, we can calculate the entropy of dS space.}

\section{de Sitter entropy}\label{Sentropy}

Numerous studies, including \cite{Akbar:2006kj, Cai:2005ra} as well as other referenced works, have extensively researched the relationship between geometry and thermodynamics in FLRW spaces. In particular, researchers have modified thermodynamic equations applicable to the dS event and AH for the non-static AH of FLRW space. This AH is distinct from the event horizon, which may not even exist, and is often considered a causal horizon linked to gravitational temperature, entropy, and surface gravity in dynamic spacetimes. References such as \cite{Bousso:2004tv, Collins:1992eca, Hayward:1997jp} support the argument that these properties also apply to cosmological horizons.

The dS horizon has thermal properties, including temperature and entropy, similar to those of the Schwarzschild event horizon, which was discovered through the utilization of Euclidean field theory techniques \cite{Gibbons:1977mu}. In their study, Gibbons and Hawking \cite{Gibbons:1977mu} evaluated the thermal bath perceived by a timelike geodesic observer in dS space while carrying a (scalar) particle detector that is restricted to a small tube surrounding the observer's worldline.

Cai and Kim \cite{Cai:2005ra} originally showed the equivalence between the Friedmann equation and the first law of thermodynamics on the AH in the framework of general relativity, as well as in the Gauss--Bonnet and Lovelock gravity theories. Later, by Akbar and Cai \cite{Akbar:2006er}, this analysis was expanded to incorporate $f(R)$ gravity and scalar-tensor gravity theories. The first law of thermodynamics was therefore derived by the use of the Friedmann equations. This connection between thermodynamics and cosmology has a significant impact on our understanding of the universe \cite{cons3, Cai2, CaiKimt}.

{In this part, let us review the Hamilton--Jacobi tunnelling proposition for BHs and the FLRW model, a topic that is likely to captivate the audience. The Hawking radiation computation on the FLRW AH was executed in the Refs. \cite{Zhu:2008hn, Jiang:2009kzr}. The calculation was replicated by the authors of Refs. \cite{CaiKimt, Medved:2002zj} using the Hamilton--Jacobi method within the Parikh--Wilczek framework, initially designed for BH horizons. In this specific scenario, the emission rate of particles under the WKB approximation signifies the probability of tunnelling for trajectories that are classically prohibited, transitioning from the interior to the exterior of the horizon. This rate of emission can be delineated as $\Gamma\sim \exp{(-2\text{Im}(I))}\simeq\exp{(-\frac{\omega}{T})}$, where $I$ indicates the Euclidean action with an imaginary part $\text{Im}(I)$, $\omega$ denotes the angular frequency of the emitted particles, and the Hawking temperature is deduced from the Boltzmann factor formula, $T =\omega/2 \text{Im}(I)$. The energy of the emitted particles, labelled as $\omega$, is defined in a manner that remains unaltered as $\omega= -K^a\nabla_aI$, where $K^a$ symbolizes the Kodama vector and the action $I$ complies with the Hamilton--Jacobi equation. Although the definition of energy is coordinate-invariant \cite{DiCriscienzo:2009kun}, it depends on the choice of time.}

In the present article, we employ the irreducible energy (mass) through the utilization of equation (\ref{new2-3}) as referenced in \cite{Mukhanov:1986me, Xiang:2004sg} to determine the temperature and entropy of the dS spacetime. This provides a simple and consistent interpretation of entropy and energy for dS spacetime. The AH's volume of dS spacetime, $V_\text{AH}$, and the MSH mass, $M_\text{AH}$, emanating \cite{Faraoni:2015ula}
\begin{equation}
    \label{mass1}
    V_\text{AH}=\frac{4\pi}{3}L_\Lambda^3,~~~M_\text{AH}^{(n)}=\frac{\Lambda}{8\pi G}V_\text{AH}=\gamma\sqrt{n}M_\text{P},
\end{equation}
where, in obtaining the last equality, we used (\ref{new2-3}), and $\gamma=\sqrt{\frac{3\pi}{2}}$. The above equation shows that the MSH mass of dS spacetime is quantized, and its spectrum is identical to the BH mass given by Eq. (\ref{H1}). 

Let us assume that Hawking radiation is emitted from the dS horizon spontaneously as the system transitions unpremeditatedly from state $n+1$ to the nearest lower state, $n$. We can then refer to the frequency of the resulting thermal radiation as
\begin{multline}\label{corr1}
    \omega_0=M_\text{AH}^{(n+1)}-M_\text{AH}^{(n)}=\\\gamma M_\text{P}\sum_{i=1}^\infty\binom{\frac{1}{2}}{i}\left(\frac{\gamma M_\text{P}}{M_\text{AH}}\right)^{2i-1}
    \frac{\gamma^2M^2_\text{P}}{2M_\text{AH}}\left(1-\frac{\gamma^2M_\text{P}^2}{4M_\text{AH}^2} \right)+\mathcal O\left(\frac{M_\text{P}}{M_\text{AH}}\right)^5.
\end{multline}
 This shows that dS space radiates with a characteristic temperature $T\propto 1/M_\text{AH}$, matching the dS Hawking temperature.

 The lifetime of a quantum system in state $M^{(n+1)}$ until its decay into the lower state $M^{(n)}$, known as the characteristic time, has been defined as $\tau_n=\omega_0/\dot M_\text{AH}$, where a dot denotes time derivative, and $\dot M_\text{AH}$ is the mass loss of the dS space due to its evaporation. Due to the fluctuations of the vacuum of the quantum fields in the vicinity of the AH, it is observed that the states' width, denoted by $W_n$, is not negligible. The estimation of the width of state $n$ can be obtained through the utilization of established methods \cite{Mukhanov:1986me} and expressed as $W_n=\beta\omega_0$, where $\beta$ is a numerical dimensionless factor.
Utilizing the uncertainty relation, $W_n\tau_n\simeq1$, and the above relations, one can easily find
\begin{equation}
    \label{Mass2}
  \dot M_\text{AH}=\frac{\beta\gamma^4}{4}\frac{M_\text{P}^4}{M^2_\text{AH}}\left(1-\frac{\gamma^2M_\text{P}^2}{8M_\text{AH}^2} \right).  
\end{equation}

{In the context of BH thermodynamics, Bekenstein \cite{Bekenstein:1998aw} utilized the above argument to draw a parallel between a BH and an atomic system by highlighting the presence of a quantum number and its associated mass level, much like in atomic structures. The phenomenon of photon emission bears resemblance to atomic transitions, showcasing similarities in the underlying processes. Bekenstein goes as far as delving into the discussion of Einstein coefficients pertaining to the spontaneous emission and absorption of a Schwarzschild BH, further solidifying the comparison to atomic behaviours. A common formula is often applied to gauge the duration of the excited state of an atomic nucleus in spectroscopy, mirroring the methods used in estimating aspects of BH dynamics. }

{At this point, we utilize the Stefan--Boltzmann law to introduce an effective temperature for AH. This serves as an extension of the concept applied in establishing the temperature of a Schwarzschild BH \cite{Xiang:2004sg, Jalalzadeh:2021gtq, Jalalzadeh:2022rxx}.
If we make the additional assumption that the source of the Hawking radiation arises from the highly blue-shifted modes just inside the AH and simultaneously consider the dS horizon as a blackbody, then the emitted power can be expressed using the well-established Stefan--Boltzmann law}
\begin{equation}
    \label{Mass3}
    \dot M_\text{AH}=\sigma A_\text{AH}T^4,
\end{equation}
where $\sigma=\pi^2/60$ is the Stefan--Boltzmann constant, and $A_\text{AH}=4\pi L_\Lambda^2$ is the surface area of the AH. Equating the mass loss terms from Eqs. (\ref{Mass2}) and (\ref{Mass3}), and selecting $\beta=\frac{\sigma}{\pi\gamma^2}=\frac{2}{45}$, we obtain the dS temperature
\begin{equation}
    \label{Mass4}
    T_\text{AH}=\frac{1}{2\pi L_\Lambda}\left( 1-\frac{\gamma^2L_\text{P}^2}{2L_\Lambda^2}\right).
\end{equation}

It should be noted that as the authors of \cite{Alicki:2023rfv} demonstrated, the temperature (\ref{Mass4}) is
only well defined in the cosmic rest frame, in which the background acts as a physical heat bath whose energy density
obeys the Stefan--Boltzmann law. This justifies the application of (\ref{Mass3}) for the AH of dS space. Furthermore, the Stefan--Boltzmann law may be used to derive the thermodynamical characteristics of a variety of BHs, as demonstrated by Refs. \cite{2004IJMPD85X,Jalalzadeh:2022rxx,Jalalzadeh:2021gtq,Hod:2016hdd,Giddings:2015uzr} and others.

In this article, we employ the first law of thermodynamics to obtain the entropy of AH. The first law of thermodynamics for AH of FLRW universe, under the
name of `unified first law,' \cite{Faraoni:2015ula} is given by
\begin{equation}
    \label{Mass5}
    T_\text{AH}\dot S_\text{AH}=\dot M_\text{AH}+\frac{p-\rho}{2}\dot V_\text{AH},
\end{equation}
where $S_\text{AH}$ is the entropy of dS AH, and $p=-\rho=\frac{\Lambda}{8\pi G}$ represent the pressure and the energy density of the vacuum in our study. Generally, $\rho$ and $p$ are the energy density and the pressure of the cosmic fluid. It is important to note that the equation mentioned above represents the differential form of the Friedmann equation \cite{Akbar:2006er}. Although the expansion law varies in different gravity theories for both flat and non-flat universes as demonstrated in Refs. \cite{Cai:2012ip,Sheykhi:2013oac}, the unified first law remains consistent in any gravity theory.  Substituting the mass and volume change rates--both of which are intrinsically negative--into the unified first law, $\dot M_\text{AH}=-\frac{\dot L_\text{AH}}{2L_\text{P}^2}$, $\dot V_\text{AH}=-4\pi L^2_\text{AH}\dot L_\text{AH}$ into unified first law (\ref{Mass5}) gives us the entropy of the AH
\begin{equation}
    \label{Mass6}
    S_\text{AH}=S_\text{dS}-\frac{5\pi\gamma^2}{2}\ln(S_\text{dS})+\text{const.},
\end{equation}
where $S_\text{dS}=\pi L_\Lambda^2/L_\text{P}^2=A_\text{AH}/(4G)$ is the well-known entropy of the dS horizon.

This shows that the entropy is quantized evenly, with the spacing between the entropy spectrum given by $\Delta S=6\pi^2$. Furthermore, the reading of $\Delta A_\text{AH}=24\pi^2$ reveals the quantum of the horizon area. The formula for entropy (\ref{Mass6}) resembles the Bekenstein--Hawking entropy formulation for BHs \cite{Ashtekar:2004nd}. However, it is important to note that in this particular instance, the horizon's properties depend on the observer, leading to ambiguities with regard to which BH concepts can be extended to dS space.

It should be noted that the same leading order logarithmic correction term has been obtained using other methods, as evidenced by Refs. \cite{Zhu:2008cg,Zhu:2009qc}. In Ref. \cite{Zhu:2008cg}, the authors showed that by applying the generalized uncertainty principle (GUP) to the AH of dS space, one can obtain the logarithmic term with a positive coefficient $\pi\alpha/4$, where $\alpha$ is a dimensionless constant realizing GUP. In Ref. \cite{Zhu:2009qc}, by applying the tunnelling method, the author obtained the logarithmic term with a positive coefficient $4\pi\alpha_1/3$, where $\alpha_1$ is an unknown constant. On the other hand, the authors of Ref. \cite{Arenas-Henriquez:2022pyh} find the same correction term with a negative unknown coefficient, by arguing that the entropy of dS space corresponds to the entanglement
between disconnected regions.
Our research has shown that the coefficient of this term is determined by the dimensionless positive constant $\gamma=\sqrt{3\pi/2}$, which is, in turn, dependent on the regularization technique utilized in Eq. (\ref{abb0}). Furthermore, incorporating higher-order correction terms from expansion (\ref{corr1}) leads to additional entropy terms proportional to the inverse powers of $S_\text{dS}$. These extra terms and the logarithmic term are typically viewed as quantum correction terms for the semiclassical analysis \cite{Ashtekar:2004nd}. The logarithmic correction completely predominates over other corrections in the large limit of the dS radius. Furthermore, since the logarithmic entropy correction is inextricably linked to the structure of all quantum gravity models, regardless of the methods used, it appears to be universal.

\section{Conclusion}\label{sec4man}
In this investigation, we conducted a rigorous thermodynamic analysis of dS space, particularly emphasizing its AH. Employing the unified first law of thermodynamics as our analytical framework, we derived an entropy formula for the dS AH incorporating quantum corrections. Our finding of a logarithmic correction term enriches the complexity of dS AH thermodynamics and challenges the conventional understanding. Significantly, we unveiled that the entropy of dS AH exhibits quantized spectral spacing, a characteristic reminiscent of black hole entropy.
However, it's crucial to acknowledge the observer-dependent attributes of the dS horizon when attempting to extend black hole thermodynamic concepts to dS spaces.
The implications of our research are manifold: they deepen the scientific comprehension of dS space thermodynamics, open avenues for exploring the intriguing but nuanced relationship between dS space and black holes, and lay a foundation for potential advancements in cosmological applications.

\ack
S.J. acknowledges financial support from the National Council for Scientific
and Technological Development -- CNPq, Grant no. 308131/2022-3.
\vspace{.3cm}
\section*{References}
\bibliographystyle{iopart-num}       
\bibliography{WDWLambda}   

\end{document}